\documentclass[11pt,letterpaper,twoside]{article}
\usepackage[margin=1.0in]{geometry}

\usepackage{graphicx}
\usepackage[font=small,labelfont=bf]{caption}
\usepackage{subcaption}
\usepackage{sidecap}
\sidecaptionvpos{figure}{t}

\usepackage{url}

\usepackage[round]{natbib}

\usepackage[dvipsnames]{xcolor}
\definecolor{grace}{rgb}{0.5, 0.3, 0.7}
\definecolor{christi}{rgb}{0.0, 0.58, 0.71}

\begin{document}

\renewcommand{\thefootnote}{\fnsymbol{footnote}}
\begin{center}
    \begin{Large}
    {\sc Astronomy in Appalachia: \\
    Five Lessons in Designing a Planetarium Show}\\
    \end{Large}
    C.~Erba\footnotemark[1]{}\footnotemark[2], 
    G.~Anderson\footnotemark[1],
    T.~Cox, G.~D.~Henson, R.~Ignace \\
    {\it Department of Physics and Astronomy, East Tennessee State University, \\ Johnson City, TN 37614, USA}
    \footnotetext[1]{co-first author}
    \footnotetext[2]{corresponding author; christi.erba@gmail.com}
\end{center}

\section*{An Ambitious Project}

``We're going to build a planetarium show!''
\vspace{1 em}

These are not exactly the words most students expect to hear in a kick-off research meeting. In our case, the students -- Grace and Trevor -- were no exception. ``Dr. Erba's enthusiasm was contagious,'' Grace remembers, "but as I glanced over at Trevor, a wave of nervousness rushed over me. Would this even be possible? Were we really going to tackle such a huge project?" 

The proposed venture, titled ``Astronomy in Appalachian Culture,'' was the development of a brand new planetarium show highlighting the connections between Appalachian folk science and the systematic description of Astronomy that is offered within college classrooms. 
The project was set to span the course of an academic year, and promised an interdisciplinary activity that would involve scholarly reading and writing, emphasize science communication, and expand the students' facility with coding. It was an ambitious undertaking for any student's Junior year research experience.

Planetariums, of course, are an invaluable resource in the science education and outreach toolbox. They offer a combined audio-visual adventure that can transport audiences around the globe, to the outer reaches of the Universe, to the deepest oceanic trenches, and back again within the span of minutes. The immersive experience captivates visitors of all ages and enables access to new locations, languages, cultures, and environments -- all without needing to leave your seat. Planetariums essentially turn the world into an educational ``sandbox," within which new ideas can be constructed, shaped, and explored.

The planetarium on our campus, East Tennessee State University, has been an important part of our school since it first opened in 1962. It is used extensively for astronomy courses taught in the Department of Physics and Astronomy, as a resource for K-12 classrooms in the region, and for community outreach through both general public and special group presentations. In addition to the hundreds of college students that experience programming in the planetarium, over a thousand visitors from off-campus also attend planetarium presentations each year. ETSU itself is nestled in Johnson City, Tennessee, about 10 minutes from the state's historic first town of Jonesboro, and squarely within the culture and traditions of Appalachia. With a campus population of over 10,000 undergraduates -- plus graduate students, faculty, staff, and administration -- the campus plays an important role in the wider community of East Tennessee.

The connection between Astronomical themes and the traditions of Appalachia have been explored extensively by ethnographers, folklorists, poets, and many others in previous works. Our goal was not to contribute discoveries to this body of knowledge, but to package the preexisting scholarship in a novel, entertaining format accessible to the general public. That said, what we learned through this project is not something that can be found in an average textbook. It required a blend of creativity, scientific knowledge, and cultural fluency to accurately reflect our different descriptions of the natural world. 
We spent countless hours researching, scripting, and producing a show that would resonate with our audience and celebrate the beauty of East Tennessee's Appalachian culture. Now, we'd like to share a few of those lessons with you.

\section*{Lesson 1: Themes Matter}

Cultural Astronomy sits at the intersection of the observation and examination of the natural world -- viewed through the perspective of science -- and descriptions of the human experience interpreted through the lenses of history, language, philosophy, and sociology.
It invites conversation around how separate cultures have perceived the Universe and our collective place within it, and explores the systems through which astronomical phenomena have been integrated into cultural traditions, creeds, and daily life. Essentially, Cultural Astronomy strives to develop a deeper understanding of the Universe in conjunction with the myriad ways humans have created meaning out of the vastness of the cosmos.

Like any show, a planetarium program requires a cohesive structure; thus developing an outline of our project that could fit within a standard 45 minute schedule became our first major task.
We drew inspiration from the common themes revealed in our deep-dive into the stories of Appalachia: the Moon, zodiac, and the seasons featured heavily in the Appalachian agricultural calendar, and strong connections between celestial objects and regional spiritual traditions emerged.
We decided to divide the show into four parts, simultaneously tracing the expanse of the cosmos illustrated against a backdrop of Earth's seasons:
\begin{enumerate}
    \item Summer: navigating the celestial sphere, the zodiac, and the Man of Signs
    \item Fall: facts and folk narratives about the Moon, Earth's nearest neighbor; the harvest season
    \item Winter: stars and planets, supernovae, and the Christmas Star; the Dead Signs of the Zodiac
    \item Spring: the Milky Way galaxy and beyond; Planting by the Signs 
\end{enumerate}
Each successive part thus represented one-quarter of Earth's trip around the Sun and one step ``outward`` into the Universe, while simultaneously staying anchored to the rhythms of life in a farming community.

Our key take-away was that the choice of scaffold is foundational to the progression of the story. Choosing our themes helped us to maintain focus in our research and provided a metric against which each new piece of information would be measured. We could proceed confidently knowing that we had deconstructed our big, idea-sized topic into a series of achievable goals. 

\section*{Lesson 2: Plant 'em in the Arms}

\renewcommand{\thefootnote}{\textcolor{black}\arabic{footnote}}

Astronomical phenomena are pervasive in the practices and oral traditions of the Appalachian regions\footnote{By definition, Appalachia spans 13 states, stretching from southern New York and Pennsylvania through parts of Ohio and Maryland, all of West Virginia, the eastern half of Kentucky and Tennessee, western Virginia and North Carolina, and the northern regions of Mississippi, Alabama, Georgia, and South Carolina. There are thus a vast number of microcultures within this geographic expanse. For our project, we focused on traditions geographically local to ETSU. For more information, see \citet{arc}.}. 
We discovered a plethora of local legends, how-to guides, moral allegories, etiologies, ghost stories, poems, and other forms of generational wisdom covering all aspects of life that included a large cast of celestial ``characters'' like the moon, stars, and planets. We were swimming in a sea of images, written media, and oral histories.

The Appalachian region's roots in farming and harvesting are fundamentally connected to Earth's seasons. The land provided rich soil and other resources to its indigenous populations, settlers, and other occupants, and Appalachia's residents used the seasons as a guide for every area of life. This included topics such as when to plant and harvest their crops, when to slaughter animals, when to can and preserve food, and when to visit the doctor. As Patrick Gainer writes in his book, {\it Witches, Ghosts and Signs: Folklore of the Southern Appalachians}, ``People who lived close to the soil learned to interpret the language of nature. This knowledge was stored in the minds of people and handed down to future generations as a great treasury of unwritten knowledge, which is our folk heritage\nocite{gainer} (p. 112).'' 
Agricultural forecasting was (and still is!) enabled by the ``Old Farmer's Almanac,'' with its dependable tables predicting the phases of the Moon and the signs of the Zodiac \citep{ofa}. Indeed, our natural satellite and these twelve constellations along the ecliptic form the basis of ``Planting by the Signs,'' a tradition that ties agricultural metrics and medical practices to the celestial dome\footnote{A useful visualization to conceptualize astrological medicine is provided by the {\it almanac man} or {\it Man of Signs}, see also \citet{ofa,hall}.}.

The title of this lesson, ``Plant 'em in the Arms,'' refers to a quote we discovered from an interview with Harriet Echols, as heard on the Foxfire podcast {\it It Still Lives}\nocite{foxfire}. Echols describes the wisdom of Planting by the Signs, saying, ``And when you sow your plants, at different signs, and when you plant your beans, the best time is from, to plant ‘em in the arms, to have a lot of beans (25:47).'' The reference to the arms is from the {\it Man of Signs}, an image mapping the Zodiac constellations onto a human body. The arms are associated with the constellation Gemini, typically signaling the planting of crops in late May and early June \citep{hall}.

This section of the project not only taught us some fantastic gardening tips, but illuminated the unexpectedly wonderful directions into which research can lead. The moments of human connection, provided through the transcribed oral histories and interviews we reviewed, served as an anchor to a topic which could easily stray to the esoteric. 
We were reminded to keep an open mind, encouraged to ask more questions, and inspired to exercise our creativity in new and exciting ways.

\section*{Lesson 3: Know Your Equipment}

Simply stated, an appreciation for the capabilities of ETSU’s planetarium equipment was crucial before we could develop the project past its initial research phase. The ETSU Planetarium was a part of the original construction of the campus's Hutcheson Hall. Its original star projector was a mechanical Spitz A3P which served the planetarium for almost 50 years, during which the facility was modestly enhanced through the improvisation and use of peripheral audio-visual devices such as slide projectors, portable video projectors, and a single channel sound system.  
With full-dome video projection technology becoming more prevalent and affordable for smaller facilities, the planetarium underwent a complete renovation from floor to ceiling in 2011. Although the original 25-foot diameter dome was kept, it was painted a flat matte gray suitable for video projection, and new theater style seating was installed in a unidirectional layout which allows for an audience capacity of 45 persons. 

Most significantly, the renovation introduced a Digitarium Epsilon (Digitalis Education Solutions, Inc.) planetarium projection system to replace the old Spitz projector. This modern, full dome digital DLP projector has a 1200-pixel resolution and a very high 2200:1 contrast ratio. The system is controlled through \texttt{Nightshade}\textsuperscript{\tiny\textregistered}\footnote{Accessible at \url{https://www.nightshadesoftware.org/projects/nightshade}.}, a simulation and visualization software platform which was developed from the open source software program \texttt{Stellarium}\footnote{Accessible at \url{https://stellarium.org/}.}. The audio output from the projection hardware is sent to a SONY digital cinema sound amplifier providing 5.1 channel output to professional quality speakers placed around the periphery of the dome. Users have the option of using a handheld remote control or a computer software interface through a web browser available on most portable devices for increased versatility in presentations.  In addition, users can create scripts to aid in the development of their own shows. These scripts can include images, sound tracks, videos, and \texttt{Nightshade}\textsuperscript{\tiny\textregistered} commands for a complete prerecorded program.

In spite of this impressive equipment, we encountered two major technological hurdles that would affect the visual and audio capacity of the final show. We learned that our particular version of \texttt{Nightshade}\textsuperscript{\tiny\textregistered}, while fully functional, was depreciated; and we found out that shorter coded scripts ran more reliably on the system. We thus decided to break the show into four ``acts,'' corresponding to each of the four seasons described above. The lack of a continuous script meant that we had to pause the show in between seasons -- but this actually turned out to be a {\it benefit} rather than a {\it detriment}. Coded scripts with a shorter length made the visuals both easier to edit and easier to pair with narration at a later stage of the project. 

We also made significant changes to our original ideas for incorporating audio tracks into the show. These decisions were partly driven by the technological challenges we faced with seamlessly operating the equipment, by limited time to fully explore the breadth of our resources, and by navigating the relatively complex world of music licensing. We ultimately made the decision to forego adding musical tracks, and to use live narration of the visual elements in the final production. The depth of the show was thus somewhat limited; however, we found creative solutions to work within our time, budget, and expertise constraints. The show must go on! 

\section*{Lesson 4: Coding (and Scripting) is an Exercise in Patience}

When we say we ``built'' a planetarium show, it's because we did: every image on the screen, every pause, every word of the script was precisely placed, timed, and tied together to create something brand new. Our show was {\it constructed}, piece by piece, from individual raw elements -- and those pieces came together in a new way when we began to code.

To anyone who has ever spent time writing code -- or writing a narrative script -- this lesson is no surprise. We had spent five preceding months involved in diligent research, but this task proved to be the most labor-intensive part of the project.
\texttt{Nightshade}\textsuperscript{\tiny\textregistered}, the software platform described above, has its own unique language structure for entering and executing commands via user-generated scripts. These formed the visual backbone of the show, to which narration or other elements could later be added.

The primary hurdle was that none of the main show creators had experience with \texttt{Nightshade}\textsuperscript{\tiny\textregistered}. Moreover, both students were relatively new to programming, with minimal previous experience using code to perform elementary calculations and create simple graphs. The learning curve was steep, but we received critical hands-on assistance from ETSU's Planetarium Director, Dr. Gary Henson, whose expertise and guidance were invaluable as we navigated the computational challenges involved in programming the show.

The final stage of production involved fitting our narrative to our images. Due to time and technological constraints, we did this the ``old-fashioned'' way, creating a visual sequence that ran on a timer with the corresponding script read live as the images traversed the planetarium's dome. On the one hand, this created a lot of headaches as we tried to adjust the visuals' timer to Grace and Trevor's individual speaking pace. On the other, this lent a unique personal element to each showing of the final program.

A word to the wise: if you create your own planetarium show, plan a little extra time into the schedule for troubleshooting code!

\section*{Lesson 5: Have Fun!}

If you're going to spend eight months (or longer) on a research project, it's important to have some fun along the way. Early in the design stage, we chose to highlight elements that were relevant both to the overarching themes of the show and to the students' personal experiences growing up in Appalachia. Grace and Trevor each chose topics that resonated with parts of their lives, adding an intangible personal significance to the program. 

Interdisciplinary projects offer a unique opportunity to explore different types of scholarly media. The initial phases of our research led us through essays and poetry, oral narratives, recorded interviews, and music that all spoke to the vibrancy and diversity of the many individual cultures that make up the Appalachian regions. We discovered the roots of popular sayings from our childhoods (``If March comes in like a lamb, it will go out like a lion; if it comes in like a lion, it will go out like a lamb;'' Gainer, p. 116), and received advice (indirectly) about all manner of issues, from planting and harvesting, to cooking and cleaning, medicine, dentistry, and child-rearing. We gained a new appreciation for how the movements of the heavens influence the rhythms of life on Earth. 

\section*{Final Thoughts}

Our completed ``Astronomy in Appalachian Culture'' program was presented twice in the ETSU Planetarium, once as a special Department of Physics and Astronomy seminar open to the University community, and once as a part of the regular public outreach schedule, open to the wider local community. The turnout was on par with other ETSU planetarium shows, and the audience feedback was overwhelmingly positive. A question-and-answer period was included at the conclusion of each show, where the audience asked detailed questions about the themes and topics included in the program. Their engagement was a compliment, highlighting the success of the program in piquing the audience's curiosity, and underscoring an overall appreciation for cultural astronomy. Their enthusiasm and encouragement reaffirmed our belief in the project's ability to educate, inspire, and connect with the audience on a deeper level.

Beyond the ETSU Planetarium, our show formed the basis of several research talks presented at local and national conferences and to the local Astronomy Club at Bays Mountain Park and Planetarium. We have benefited from the opportunity to speak to a wide range of audiences, both about how astronomy and Appalachian culture are integrally linked, and about the lessons we learned in undertaking such an ambitious endeavor. We have taught, and we have been taught by, our discoveries throughout this interdisciplinary project, and we have been motivated to further explore innovative ways to educate and relate to our community. As we look to the future, we are excited to expand upon this foundation, and to ensure that the show continues to inspire and resonate with audiences well beyond the edge of the dome. 



\begin{figure}[htpb]
     \centering
     \begin{subfigure}[b]{0.3\textwidth}
         \centering
         \includegraphics[width=\textwidth]{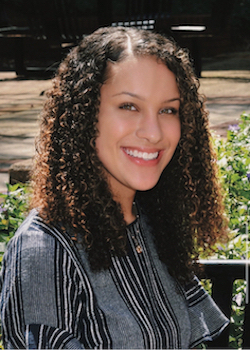}
         \label{fig:anderson}
     \end{subfigure}
     \hfill
     \begin{subfigure}[b]{0.3\textwidth}
         \centering
         \includegraphics[width=\textwidth]{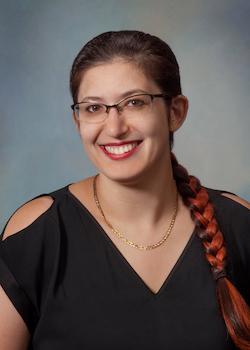}
         \label{fig:erba}
     \end{subfigure}
     \hfill
     \begin{subfigure}[b]{0.3\textwidth}
         \centering
         \includegraphics[width=\textwidth]{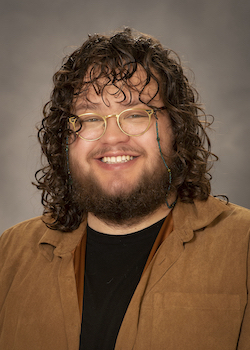}
         \label{fig:cox}
     \end{subfigure}
        \caption{Article authors Grace Anderson (left), Dr. Christi Erba (center), and Trevor Cox (right), the creators of the ``Astronomy in Appalachian Culture'' Planetarium show.}
        \label{fig:authorphotos}
\end{figure}

\section*{Acknowledgements}

G.A., C.E., and R.I. gratefully acknowledge support for this work from the National Science Foundation under Grant No. AST-2009412. G.A. and T.C. gratefully acknowledge support from the ETSU Ronald E. McNair Postbaccalaureate Achievement Program, a TRIO program that sponsored G.A.'s and T.C.'s participation in the Advanced Research Internship. The authors also thank Dr. Ted Olson, from ETSU's Department of Appalachian Studies, for his guidance and direction in the early stages of this project. 

\bibliographystyle{chicago}
\bibliography{ref}

\end{document}